# Extragalactic Planetary Nebulae (xPNe) – Determining Distances out to 100 Mpc and the Renaissance of the PN Luminosity Function Method

Magda Arnaboldi[1], Ortwin Gerhard[2], Martin Roth[3], Peter M. Weilbacher[3], Souradeep Bhattacharya[4], Johanna Hartke[5], Chiara Spiniello[1], Azlizan Soemitro[3], Claudia Pulsoni[2], Lucas Valenzuela[6], George Jacoby[7], Robin Ciardullo[8]

[1] ESO, Garching, Germany
[2] MPE, Garching, Germany
[3] AIP, Potsdam, Germany
[4] CAR, Univ. Hertfordshire, UK
[5] FINCA, Turku, Finland
[6] LMU, Germany
[7] NSF's NOIRLab, USA
[8] Department of Astronomy & Astrophysics, The Pennsylvania State University, USA

**The discrepancy of the Hubble parameter H0 as measured from the cosmic microwave background versus that found from traditional distance ladder measurements has produced considerable discussion about the need for another force in cosmology. However the significance of the discrepancy depends on understanding the systematics associated with crowding, metallicity effects, and extinction of the stellar tracers. Thus additional precision distance indicators in the local universe are desperately needed for investigating the H0 tension. The analysis of MUSE archival data makes the case that the Planetary Nebula Luminosity Function (PNLF) has become such an indicator, as the method can reach distances comparable to HST distances of Cepheids at a fraction of a cost, in terms of telescope time and ground-based. With new wide-field spectroscopic facilities it becomes possible to measure distances to early-type galaxies (ETGs) using the PNLF out to 100 Mpc distance, achieving a precise estimate for the H0 value which is independent of the Type Ia supernova calibration, with only single-epoch measurements.**

## Introduction

The technique of obtaining distances using the [OIII] 5007Å planetary nebula luminosity function (PNLF) is now over three decades old (Ciardullo+1989). Today, observations through wide-field integral field spectrographs (IFS) have supplanted interference filter-based photometry for deep PNLF observations. In particular, the Multi-Unit Spectroscopic Explorer (MUSE) optical IFS (Bacon+2010) has revolutionized PNLF measurements to both spiral and elliptical galaxies by coupling the one arcmin$^2$ aperture and excellent image quality of the 8.2 m Very Large Telescope (VLT) with the resolution of an R ~ 2000 spectrograph. By reducing the sky and galaxy noise underlying each PN by more than a factor of 10 and allowing the simultaneous measurement of several spectral lines (such as [O III] λ5007, Hα, and [S II] λλ6716, 6731), MUSE has facilitated the identification of large numbers of extragalactic planetary nebulae (xPNe) with almost no contamination from interloping objects, such as H II regions, supernova remnants (SNRs), and background emission-line galaxies. The result has been robust PNLF distance measurements to dozen spiral and elliptical galaxies within ~30 Mpc (Jacoby+2024; hereafter J+2024 and references therein). By using MUSE archival data for 16 galaxies, two of which are isolated and beyond 30 Mpc in a relatively unperturbed Hubble flow, J+2024 derive a Hubble constant value which is in a competitive range with respect to those determined before e.g., using Cepheids, the tip of the red giant branch (TRGB), and surface brightness fluctuations (SBF). The J+2024 study was limited by the fact that the observations were not taken for precise distance determination purposes: if they were, the errors would have been smaller.

## Why is it that using the xPNe as distance indicators is compelling and why now?

The tension between Hubble constant (H0) values derived from distance ladders (~73 km s$^{-1}$ Mpc$^{-1}$; Freedman+2020; Riess+ 2022,2024; Freedman & Madore 2023) and H0 derived from early universe measurements (~67 km s$^{-1}$ Mpc$^{-1}$; Planck coll. 2020) has reached a level of 5 sigma in the Cepheid-based determinations, potentially indicating new physics. *This claim of new physics must be investigated using as robust evidence as possible, in as many different ways as possible.* The PNLF is an established secondary distance indicator (Ciardullo 2012) which works by comparing the PNLFs measured in any galaxy with an analytic template function (Ciardullo+ 1989) calibrated from galaxies with known distances using a variety of methods (Cepheids, TRGB, SBF and geometric means; see Ciardullo 2022). The absolute magnitude PNLF reproduced from J+2024 with M*=-4.54 is shown in Figure 1. The shift in magnitude from the galaxy measured PNLF bright cut-off to the absolute magnitude PNLF gives directly the distance modulus. While the reliability of the PNLF has been questioned in the past due to the lack of a robust theory for the phenomenon, new studies are changing this (Valenzuela+2025, Soemitro+2025 (hereafter S+2025), Jacoby & Ciardullo 2025).

The PNLF distance determinations have been greatly improved since the classical narrow band *on-off* imaging (Arnaboldi+2002) by the advent of *MUSE with ground-layer adaptive optics (AO) correction at the VLT* (Kreckel+2017; Fensch+2020; Spriggs+2021; Scheuermann+2022; Congiu+2025). Additionally, Roth+2021 developed an optimized data analysis technique labelled ``differential emission-line filter'' (DELF hereafter), which increased the effectiveness of PN detections by another 25%. Using MUSE archival data for several galaxies, J+2024 showed that, when the image quality is excellent (FWHM≤0".75), the DELF analysis of the MUSE data cubes can deliver precise (≤ 0.07 mag) [O III] photometry for PNe at distances out to ~30 Mpc. The combination of MUSE's narrow effective bandpass (5 times smaller than that for on-off imaging), full spectral coverage, R ~ 2000 spectral resolution, and excellent image quality allows xPNe to be securely identified at distances out to 30 Mpc with the VLT. In other words, the PNLF can now obtain precision distances to galaxies in a regime where the galaxy peculiar velocities are only 10% that of the Hubble flow (Tonry+2000), while avoiding the issues that can lead to unwanted systematics in their distance scale (Ferrarese+2000).

Recent theoretical models (Valenzuela+2025, S+2025) showed that in simulated early-type galaxies (ETGs) with luminosity L≥10$^{10}$ L$_\odot$ no systematic dependence of M* with galaxy age (9<Age<13.5 Gyr) and metallicity (-0.4< [M/H] <+0.2) is expected. This overcomes a decade-old problem (Marigo+2004) in reconciling PNLF observations and theory, strengthening the use of the PNLF as a distance indicator. The models cannot yet predict precise values of M*, therefore M* is currently determined empirically from galaxies with other available distance measurements. This results in M* = -4.54±0.05 (Ciardullo 2012, 2022). Current work is on-going to improve this; as the Local Group (M31) distances improve, so does the PNLF zero point.

Stellar populations of the above age and metallicity ranges **dominate the light in ETGs within one effective radius**, i.e. the area which would be sampled by an IFU FOV of few arcmin$^2$ when pointing at the galaxy's centre. The green squares on the galaxy images in Figure 1 show the pointings of a FOV of one arcmin$^2$ (conservative hypothesis) of a future IFU facility with R~2000 on a 10+meter telescope. The area covered will allow the detection of 50-100 xPNe in each object for reliable PNLF distance determinations. The combined recent results from VLT-MUSE archival data, theoretical models and simulations make the case that the PNLF technique with a new IFS facility with a FOV>1 arcmin$^2$, R~ 2000, on a 10+ meter class telescope ideally equipped with ground-layer AO to boost the detection of point-like sources may become a transformative tool for obtaining precise extragalactic distances _beyond_ 30 Mpc. Very important is that the method can be applied to independent tracer galaxies, ETGs, distinct from those usually utilized for Cepheids/TRGB distance determinations.

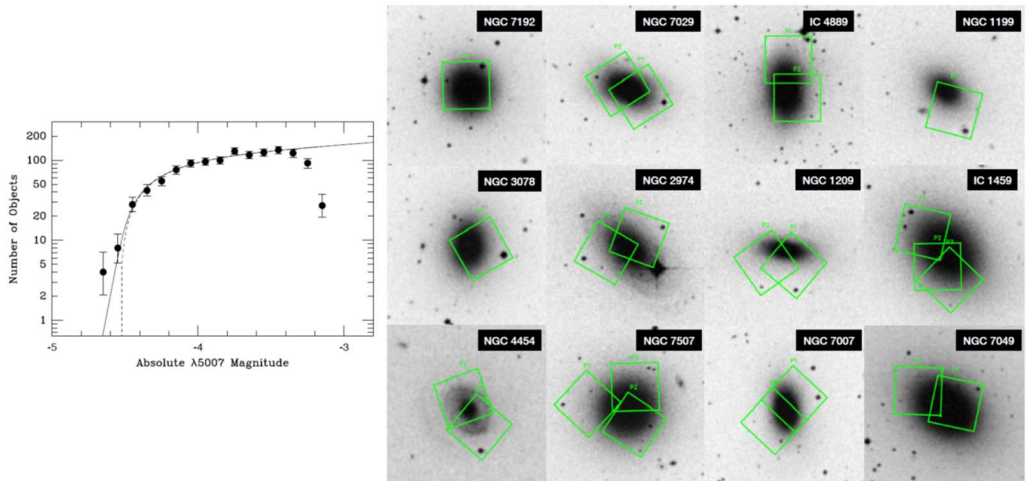

Figure 1: Left panel: The combined PNLF from 1016 PNe in the complete samples of 14 high-quality galaxies in J+2024. The PNe's absolute magnitudes have been computed using each system's most likely apparent distance modulus. The dashed line shows the assumed empirical PNLF (Equation (2) from Ciardullo+1989); the solid curve gives the expected PNLF when accounting for the formal errors of each galaxy's distance determination as applied to the assumed empirical PNLF. Right panel: images of selected galaxies for a possible ETGs PNLF sample from DSS. North is up and East to the left. The green squares in each image show the pointings of a FOV of 1 arcmin$^2$ (conservative hypothesis) of a future IFU facility on a 10+meter telescope. The area covered will allow the detection of 50-100 xPNe in each object for reliable PNLF distance determination.

## A new facility to reach detection of xPNe out to 100 Mpc

For a distance modulus of 34.9, D=100 Mpc, a xPN at the PNLF bright cut-off has a [O III] 5007 Å line flux of $2 \times 10^{-18}$ ergs s$^{-1}$cm$^{-2}$. To be able to sample the PNLF cut-off in an ETG in the Coma cluster, one would need to reach a limiting flux of ~$1 \times 10^{-18}$ ergs s$^{-1}$cm$^{-2}$ as shown by the ground breaking results of Gerhard+2005,2007. When these faint fluxes can be detected with such a new facility, they allow the detection of O(50) xPNe in any galaxy of with total luminosity L≥$10^{10}$ L$_\odot$ out to 100 Mpc distances. Surveying O(20) galaxies would give us an accuracy <1% on the H0 value. Given a sufficient number of such independent xPNe tracers in galaxies, covering the 30-100 Mpc distance range and regions with small peculiar velocities, the PNLF technique with a new IFS facility on a 10+ meter telescope ideally equipped with ground-layer AO can ``_place a meaningful constraint on the Hubble constant within our backyard that is independent of the Type Ia supernovae calibration_" (J+2024, Scolnic+2025) using only ground based facilities.

## References

Arnaboldi+2002AJ....123..760A; Bacon+2010SPIE.7735E..08B; Ciardullo+1989ApJ...339...53C; Ciardullo2012 Ap&SS.341..151C; Ciardullo2022 FrASS...9.6326C; Congiu+2025A&A...700A.125C; Ferrarese+2000 ApJS..128..431F; Fensch+2020A&A...644A.164F; Freedman+2020ApJ...891...57F; Freedman & Madore 2023 JCAP...11..050F; Gerhard+2005ApJ...621L..93G; Gerhard+2007A&A...468..815G; Jacoby+2024ApJS..271...40J (J+2024); Jacoby & Ciardullo 2025ApJ...983..129J; Kreckel+2017ApJ...834..174K; Marigo+2004A&A...423..995M; Planck Coll.+ 2020A&A...641A...6P; Riess+2022ApJ...934L...7R; Riess+2024 ApJ...977..120R; Roth+2021 ApJ...916...21R; Scheuermann+ 2022MNRAS.511.6087S; Scolnic+ 2025 ApJ...979L...9S; Soemitro+2025 arxiv:2509.10175v1 (S+2025); Spriggs+2021A&A...653A.167S; Valenzuela+2025 A&A...699A.371V; Tonry+2000 ApJ...530..625T